\documentclass[nofootinbib,floatfix,longbibliography,twocolumn]{revtex4-2}
\usepackage{amsmath}
\usepackage{amssymb}
\usepackage{graphicx}
\usepackage{verbatim}
\usepackage{color}
\usepackage{paralist}
\usepackage{lipsum}
\usepackage{fancyhdr} 
\usepackage{upgreek}
\usepackage{multirow,array,hhline,booktabs} 

\AtBeginDocument{}
\usepackage[pdftex, breaklinks, colorlinks, citecolor=blue, urlcolor=blue]{hyperref}

\begin{document}
\author{Edward H. Chen}
\altaffiliation{Present address: IBM Almaden Research Center, 650 Harry Road, San Jose, California 95120, USA}
\author{Kate Raach}
\author{Andrew Pan}
\author{Andrey A. Kiselev}
\author{Edwin Acuna}
\author{Jacob Z. Blumoff}
\author{Teresa Brecht}
\author{Maxwell D. Choi}
\author{Wonill Ha}
\author{Daniel R. Hulbert}
\author{Michael P. Jura}
\author{Tyler E. Keating}
\author{Ramsey Noah}
\author{Bo Sun}
\author{Bryan J. Thomas}
\author{Matthew G. Borselli}
\author{C.A.C. Jackson}
\author{Matthew T. Rakher}
\author{Richard S. Ross} 
\affiliation{HRL Laboratories, LLC, 3011 Malibu Canyon Road, Malibu, California 90265, USA}

\begin{abstract}
Silicon quantum dot qubits must contend with low-lying valley excited states which are sensitive functions of the quantum well heterostructure and disorder; quantifying and maximizing the energies of these states are critical to improving device performance. 
We describe a spectroscopic method for probing excited states in isolated Si/SiGe double quantum dots using standard baseband pulsing techniques, easing the extraction of energy spectra in multiple-dot devices. 
We use this method to measure dozens of valley excited state energies spanning multiple wafers, quantum dots, and orbital states, crucial for evaluating the dependence of valley splitting on quantum well width and other epitaxial conditions. 
Our results suggest that narrower wells can be beneficial for improving valley splittings, but this effect can be confounded by variations in growth and fabrication conditions. 
These results underscore the importance of valley splitting measurements for guiding the development of Si qubits.

\end{abstract}

\title{Detuning Axis Pulsed Spectroscopy of Valley-Orbital States in Si/SiGe Quantum Dots}
\date{\today}
\maketitle

\section{Introduction}\label{main:intro}

Electrostatically confined quantum dot (QD) spin qubits in Si/SiGe quantum wells (QWs) are a promising platform for processing quantum information~\cite{vandersypen2019quantum,LADD2018}. 
The valleys inherent to the Si band structure, however, can fundamentally limit the initialization, manipulation, and measurement fidelities of qubit states~\cite{Zwanenburg2013}. 
In a typical Si/SiGe QW design, tensile strain and spatial quantization partially break the six-fold degeneracy of the bulk Si conduction band, leaving the two lowest out-of-plane valleys. 
This remaining degeneracy can be lifted by sharp disruptions of the periodic crystal potential at the QW interfaces. 
Measuring, understanding, and increasing the resulting valley splitting (VS) energy is critical for improving qubit performance. 

\begin{figure}[th!]
\includegraphics[width=1\linewidth]{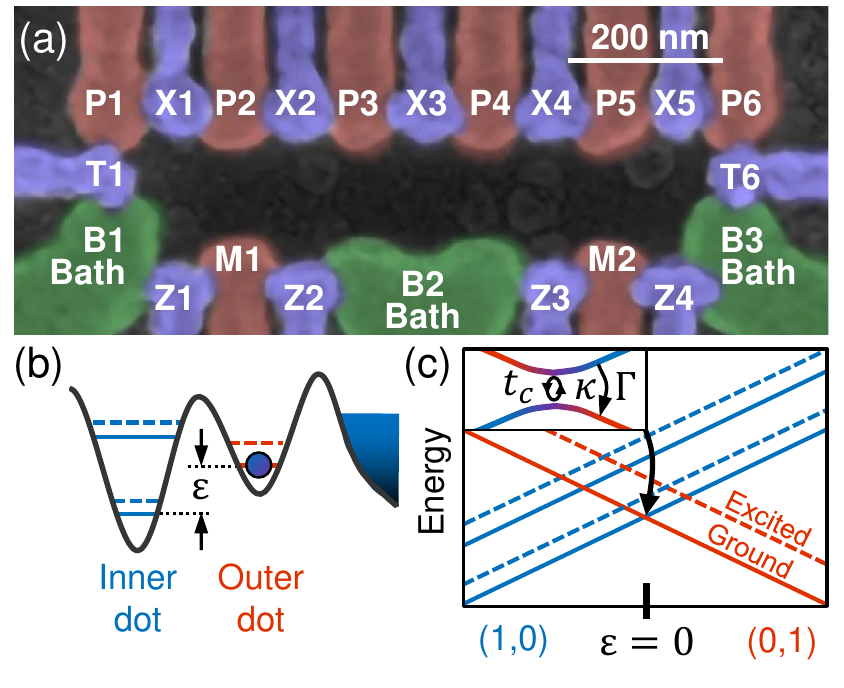}
\caption{
\textbf{(a)} Scanning electron micrograph of a six QD accumulation-mode device layout typical of those used in this study. 
Readout is done via the M1 and M2 QDs biased to the Coulomb blockade regime to measure the charge configuration.
\textbf{(b)} Schematic of low-energy excited states of a DQD; each orbital excitation (solid lines) has an associated valley excitation (dashed lines). 
The detuning energy $\varepsilon$ is the primary degree of freedom for extracting energy splittings in the DAPS technique.
\textbf{(c)} Illustration of the DQD energy level diagram along the detuning axis. 
The inset shows the vicinity of a level anticrossing, where the charge transfer dynamics are strongly impacted by the tunnel coupling $t_c$ and inter-dot dephasing and relaxation rates, $\kappa$ and $\Gamma$.
}
\label{fig1}
\end{figure}

Despite its importance, there is not a consensus on the best method of measuring VS in silicon QDs. 
Shubnikov-de Haas measurements are commonly used to extract the energy spectrum, including valley states, of 2-D electron gases induced in QWs~\cite{goswami2007controllable,sasaki2009well,neyens2018critical,wuetz2020effect}. 
However, these macroscopic ensemble measurements average over epitaxial disorder, magnetic confinement, and many-body effects, and hence their quantitative relevance to individual single-electron QDs is not clear. 

A number of techniques for directly probing QD energy levels have been demonstrated with assorted advantages and limitations. 
Measurements of QD conductance can yield information about level spacings \cite{fuechsle2010spectroscopy}, but are complicated for multi-QD devices not designed for transport. 
RF-based techniques, such as photon-assisted tunneling or resonant cavity response of double quantum dots (DQDs), can drive intervalley transitions precisely, but are typically limited by the maximum frequency of microwave photons which can be reliably delivered to the device~\cite{wang2013charge,Mi2017}. 
Pulsed-gate spectroscopy, by contrast, does not require microwave control and has been used to probe valley and other excited QD states in a variety of systems \cite{elzerman2004excited}. 
This can be used in concert with large external magnetic fields to perform magnetospectroscopy (measuring valley-limited two-electron singlet-triplet splittings, for instance) \cite{borselli2011measurement} or to extract single-electron VS via spin-valley hot-spot effects \cite{hollmann2019large}. 
However, conventional QD pulsed-gate techniques \cite{harbusch2010rf,huebl2010electron} rely on the detuning dependence of the tunneling rate with a neighboring electron bath, similar to transport-based measurements. 
This restricts the range of allowable QD biasing and can complicate the spectral interpretation when weakly quantized bath excitations induce additional resonant features~\cite{ fuechsle2010spectroscopy,simmons2011tunable}. 
Overall, commonly used techniques require devices to be designed or tuned in ways that vary significantly from ideal coherent operation, making it harder to characterize the dot-to-dot and device-to-device behavior of VS in extensible designs.

Nonetheless, using the methods mentioned above as well as others \cite{HRLJones2019,schoenfield2017coherent,penthorn2019experimental}, various groups have reported a wide range of valley splittings between 10 and 270~$\upmu$eV in Si/SiGe QDs.
The large variation is commonly assumed to be related to the QW epitaxy, though further insight is difficult as values are typically reported for just one or two QDs.

Valley splitting is sensitive to the overlap of the confined electron wave function with the Si/SiGe epitaxial hetero-interfaces, and is therefore expected to improve with better interfaces, narrower quantum wells, and/or stronger electric fields~\cite{boykin2004valleyB,boykin2004valleyA,friesen2007valley,zhang2013genetic,zimmerman2017valley}. 
Similarly, interfacial disorder originating from steps, atomic inter-diffusion, intrinsic alloy randomness, and other sources is expected to affect valley splitting energies realized by electrons in different orbital states~\cite{kharche2007valley,jiang2012effects}. 
Disorder also introduces valley-orbit mixing~\cite{gamble2013disorder} responsible for coupling valley states belonging to different orbitals. 
Thus, strictly speaking, the resulting electron eigenstates are commonly mixed, i.e., valley-orbit in character. 
Both measurements and modeling of our devices suggest the orbital energies are on the 1~meV scale whereas the valley energies are on the 10-100~ueV scale. 
When the energetics of orbital confinement strongly dominate valley mixing, it becomes appropriate to distinguish \textit{mostly orbital} and \textit{mostly valley} states and excitations, with an unambiguous structure of low-energy levels~\cite{Zwanenburg2013} emerging as a result. 
Reliable valley splitting measurements across a range of QDs and devices are necessary for disentangling these effects.

In this paper, we introduce an experimental technique for measuring the excited state energy spectra of pairs of QDs. 
This method, which we call detuning axis pulsed spectroscopy (DAPS), can be applied to any charge configuration, but we primarily focus on the single-electron case for which the spectra are most easily understood. 
This allows us to systematically probe multiple valley and orbital states of individual dots in Si/SiGe QWs with varied widths and growth conditions. 
Our results validate the scalability of the DAPS technique and emphasize the importance of epitaxial uniformity for controlling the valley splitting in QD devices.

\section{Experimental Setup}\label{main:setup}

Our experiments use single electrons trapped within arrays of accumulation-mode, Si/SiGe QDs designed to form a pair of exchange-only qubits~\cite{divincenzo2000,reed2016,HRLAndrews2019}. 
An example of such a device in an overlapping Al gate layout \cite{Borselli2015} is shown in Fig.~\ref{fig1}(a). 
The quantum dots form within an isotopically purified (800 ppm) Si QW embedded in a strain-relaxed, undoped SiGe alloy. 
Large field gates (not visible) surrounding the quantum dot region prevent the accumulation of electrons around the device perimeter. 
Six plunger gates labeled as P1 through P6 control the chemical potential within each dot, while five exchange gates labeled as X1 through X5 modulate the inter-dot tunneling rates and exchange interactions. 
Two additional tunneling gates labeled as T1 and T6 control the tunneling rates of electrons into dots P1 and P6 from reservoirs B1 and B3, respectively. 
Two dot charge sensors (DCS) formed underneath gates M1 and M2 are used for measuring charge configurations of the six-dot array. 
For coherent operations, the six electrons comprise two triple-quantum-dot exchange-only qubits, which can be fully controlled by only baseband voltage pulses and do not require external magnetic fields or RF control.
It is therefore advantageous to characterize the level spectrum in the same way, which is one technical advantage of our approach.

In devices like the one pictured here, most QDs are far away from a neighboring electron bath. 
When we characterize a DQD pair, therefore, we generically refer to the QD closer to the nearest electron bath as the ``outer'' dot and the more distant QD as the ``inner'' dot, as sketched in Fig.~\ref{fig1}(b). 
Here we also schematically depict the simplest case for QD low-energy states with each orbital (solid lines) having an associated valley excitation (dashed lines). 
The DQD energy spectrum as a function of dot detuning resembles Fig.~\ref{fig1}(c), and our measurement method relies on mapping the anti-crossings between excited levels of the different dots.

\begin{figure}[th!]
\includegraphics[width=0.8\linewidth]{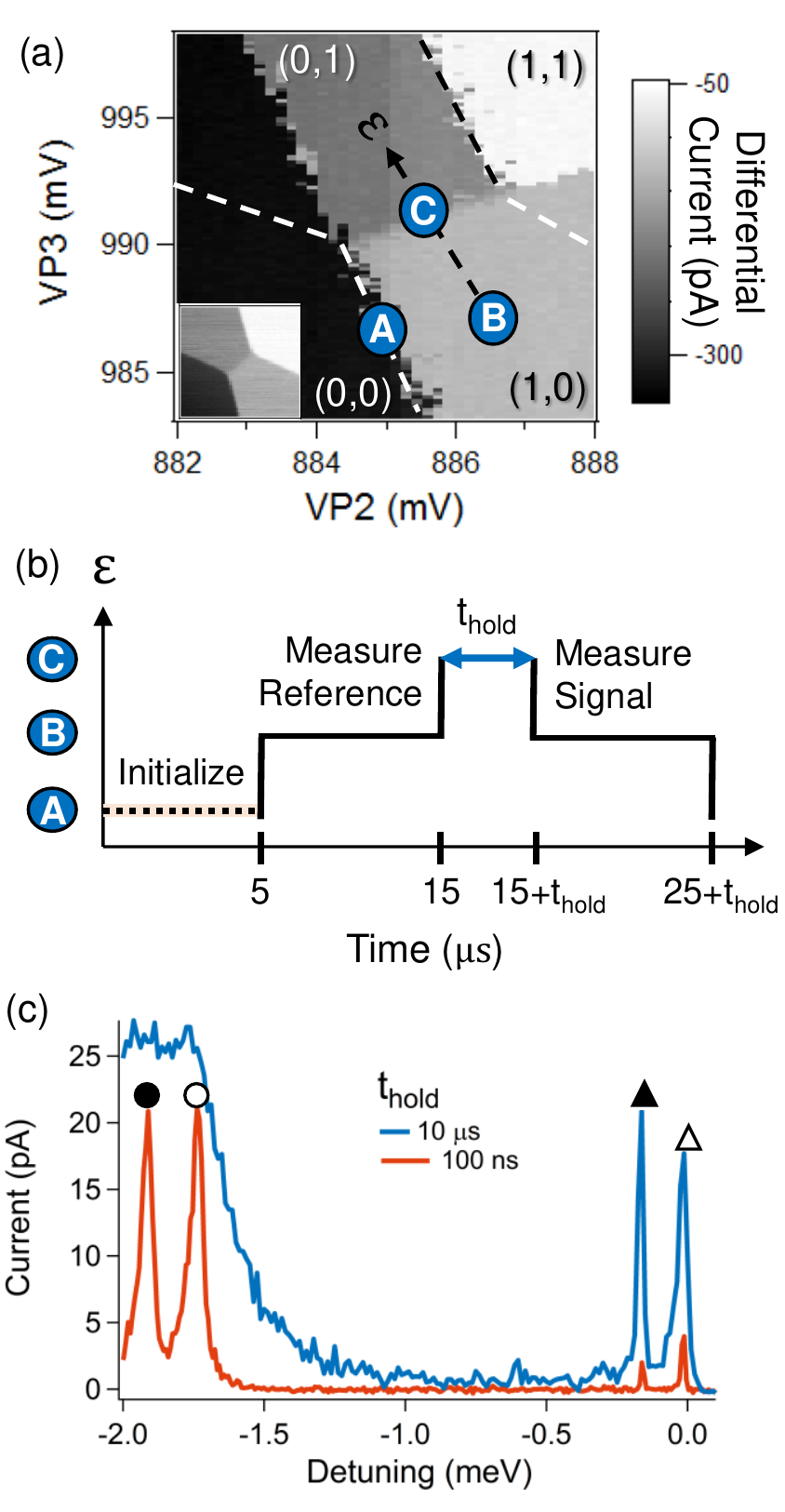}
\caption{
\textbf{(a)} Measured charge stability diagram of a P2-P3 DQD in the low tunnel coupled regime. 
Relevant bias locations for the DAPS measurement are overlaid; the line along B-C defines the detuning axis parameterized by $\varepsilon$.
The inset shows a typical charge stability diagram taken at higher tunnel coupling, which has the expected honeycomb charge cell shape. 
\textbf{(b)} Pulse timing sequence for DAPS.  
A 5~$\upmu$s initialization at A is followed by a 10~$\upmu$s charge reference measurement at B. 
Then the bias moves to the swept position C for a duration $t_{\text{hold}}$ before a final charge measurement at B lasting for 10~$\upmu$s.
\textbf{(c)} P3 DAPS measurement of differential current versus detuning energy (inferred from voltages using the measured capacitance matrix) for two hold times: 100~ns (orange) and 10~$\upmu$s (blue). 
}
\label{fig2}
\end{figure}

\begin{figure*}[th!]
\includegraphics[width=0.75\linewidth]{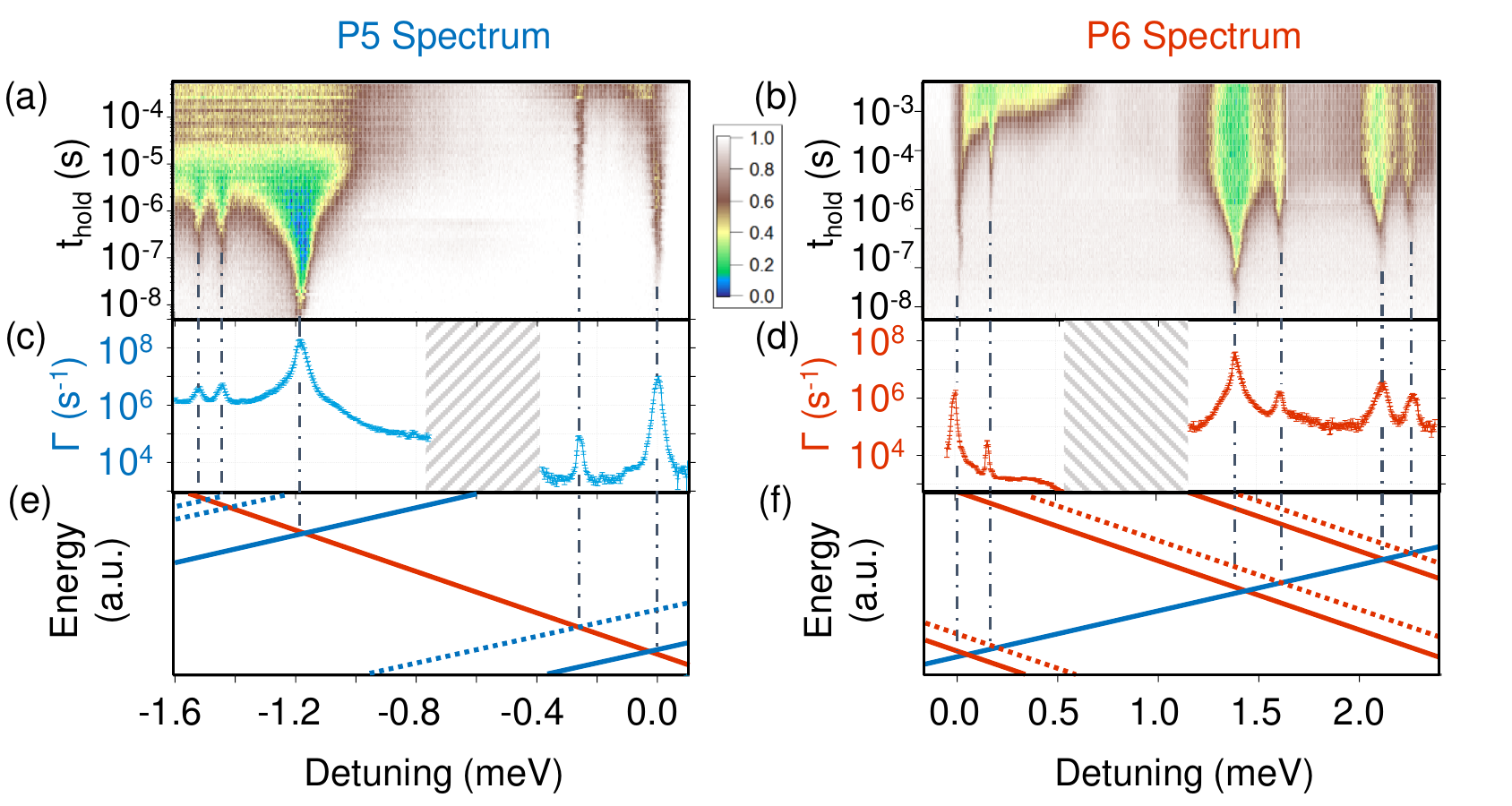}
\caption{
DAPS charge decay measurements of a P5-P6 DQD as a function of detuning $\varepsilon$ and hold time $t_{\text{hold}}$ for \textbf{(a)} the inner dot P5 and \textbf{(b)} the outer dot P6. 
The colorscale indicates probability of the electron remaining in its initial charge state.
\textbf{(c,d)} Charge decay rates $\Gamma$ vs detuning extracted from exponential fits of each column of the data in \textbf{(a,b)}.
\textbf{(e,f)} Simplified energy level diagrams vs detuning for the P5 (blue lines) and P6 (red lines) QDs. 
Notably, for each orbital in a dot, the valley ground state (solid line) tends to decay faster than its valley excited counterpart (dotted line).
A possible exception is the set of higher excited P5 orbital/valley states in the upper left side of \textbf{(e)}, as further discussed in the text.
}
\label{fig3}
\end{figure*}

\section{DAPS Technique}\label{main:daps}

The DAPS technique probes the DQD energy spectra by preparing electrons in the ground state of one dot and tracking charge transitions to the other dot as a function of detuning across the (1,0)-(0,1) charge boundary. 
Energies are extracted by identifying level anti-crossings at particular detunings, as indicated in Fig.~\ref{fig1}(c), based on enhanced charge transition rates at those biases. 
This requires only standard baseband control, without the need for RF or magnetic fields, and improves on standard pulsed-gate spectroscopy~\cite{elzerman2004excited} by probing multiple dots without requiring direct coupling to an electron bath. 
We illustrate the experimental procedure in Fig.~\ref{fig2}(a)-(b) using the (1,0)-(0,1) transition of a DQD, formed in this case underneath gates P2 and P3 in a device like that in Fig.~\ref{fig1}(a). 
Fig.~\ref{fig2}(a) is the charge stability diagram for the DQD where the M1 DCS conductance is measured while the voltage biases applied to the P2 and P3 gates are swept. 
The dots are in the low tunnel coupled regime we use to perform DAPS, where the loading rates are slower than our scan rate.
The expected charge cell boundaries become clearer at higher tunnel coupling as illustrated in the inset. 
The charge stability diagram is labeled with the three bias points A, B, and C we cycle between for the experiment. 
At point A near the (1,0) loading line, we prepare the electron in the ground state of the outer dot (here, P2); this can be done by waiting for thermalization within the charge cell, though we accelerate this process by allowing electron exchange with a neighboring bath extended by biasing the P1 QD to the same chemical potential as bath B1~\cite{maune2012}. 
The electron is then pulsed to point B where the nearest dot charge sensor (DCS), either M1 or M2, is measured to record a reference conductance for the (1,0) charge state. 
A diabatic\footnote{Adiabatic and diabatic pulses refer to the two extreme conditions of Landau-Zener transitions~\protect\cite{schoenfield2017coherent,mi2018}.} pulse takes the electron, its state unchanged, to point C along the detuning axis, $\varepsilon$, across the charge boundary, where it is held for time $t_{\text{hold}}$ before returning rapidly to B for another charge state measurement.
This second measurement is compared with the reference to obtain a differential signal.

Sweeping the detuning position C and $t_{\text{hold}}$ allows us to map out the charge decay dynamics, as shown in Fig.~\ref{fig2}(c). 
Two orbital excited states (empty circle: ground valley; filled circle: excited valley) are resolved as peaks for $t_{\text{hold}}=100$~ns, but become indistinguishable for $t_{\text{hold}}=10~\upmu$s due to rapid inter-dot relaxation at large detuning.
In contrast, the two orbital ground states (empty triangle: ground valley; filled triangle: excited valley) remain well-resolved for both hold times.
These four transitions resemble states of the inner dot illustrated in Fig.~\ref{fig1}(b).
The detuning voltage is translated to energy using the measured gate lever arm matrix~\cite{simmons2011tunable,HRLJones2019}.
Experimental uncertainties in the lever arm measurements constitute a primary source of systematic error; for devices that show microwave response, auxiliary RF measurements can reduce this uncertainty by providing a definite voltage-to-energy scale, as illustrated in App.~\ref{app:rfdaps}.
The procedure may be extended to probe the other, outer dot (P2 in this example) by preparing the electron at A and adiabatically ramping across the charge boundary to the (0,1) charge cell, from where diabatic detuning excursions towards (1,0) can be used to determine the spectrum of the P2 dot. 

The DAPS sequence resembles standard charge qubit control techniques and its main features can be described by a Lindblad master equation (analyzed in detail in App.~\ref{app:model}) where the most relevant quantities are the interdot tunnel coupling $t_c$, charge dephasing $\kappa$, and the interdot inelastic relaxation rate $\Gamma$, as sketched in Fig.~\ref{fig1}(c). 
However, coherent charge manipulation requires large tunnel coupling ($\gg$ 1~GHz) and hence ultrafast control. 
While non-adiabatic pulses can coherently probe excited states using Landau-Zener interferometry~\cite{schoenfield2017coherent,mi2018}, its extreme sensitivity to tunnel coupling and the pulse waveform makes it an unwieldy method to optimize for spectroscopy. 
Instead, we find the low tunnel coupling regime to be most desirable for DAPS because it maintains the diabaticity of detuning pulses, ensuring all charge transitions occur during the hold step (point C in Fig.~\ref{fig2}(b)). 
In this incoherent regime, when the DQD is detuned at an inter-dot level crossing, charge transitions occur at a rate approximately equal to $t_c^2/\kappa$ where both $t_c$ and $\kappa$ can range from $10^7$---$10^{10}$ s$^{-1}$. 
These longer time scales are easily probed using typical control hardware for exchange-only operation, which requires bandwidths in the 0.1-1.0~GHz range~\cite{reed2016}. 
To resolve these transitions, hold times should be kept short to prevent inelastic decay ($\Gamma\approx 10^3$-$10^8$ s$^{-1}$) from equilibrating the DQD. 
The measurement steps at point B must also be done faster than $\Gamma$ to resolve the signal. 
This implies that DAPS should be performed for $\kappa t_c^{-2}<t_{\text{hold}}<\Gamma^{-1}$. 
Meeting these criteria requires tuning both the tunnel barrier and $t_{\text{hold}}$, since $t_c$ varies for excited states and the inelastic decay rate may itself be a function of detuning and tunnel coupling. 
As one example, the excited orbital states at large detuning in the left-hand side of Fig.~\ref{fig2}(c) are only clear at short $t_{\text{hold}}$ and are unresolved at longer times due to their faster thermalization rates.

\section{Valley-Orbital Spectroscopy Results}
Using the DAPS method, we can quantify the valley-orbit level structure of Si/SiGe QDs in more detail than is usually reported. 
As an example, Fig.~\ref{fig3}(a,b) depicts the measured charge transition probabilities of a P5-P6 DQD versus detuning energy and hold time. 
We obtain the probability by averaging over single-shot DAPS measurements at each detuning and $t_{\text{hold}}$, where a charge transition is registered if the differential current between the signal and reference measurements surpasses a specified threshold. 
Since the (1,0) and (0,1) charge states have substantially different conductance signals, the binning of charge states can be performed accurately using standard QD readout procedures~\cite{HRLJones2019}. 
Resonant peaks are identified in Fig.~\ref{fig3}(c,d) from fitted exponential decay rates as a function of detuning, which we then associate with the level crossings illustrated in Fig.~\ref{fig3}(e,f). 
Table~\ref{tab:DotP5P6} lists the extracted energies and tentative assignments to mostly valley or mostly orbital states based on the splitting magnitudes and transition strengths. 
In particular, the lowest excited-valley transitions exhibit decay times roughly ten times longer than the ground-valley transitions in Fig.~\ref{fig3}(c,d), suggestive of a 3x or greater reduction of the inter-valley tunnel coupling due to differences in the inter-dot valley phase~\cite{dassarma2010}. 
This pattern also recurs for several higher excited states in both P5 and P6 dots, further motivating our valley-orbit assignments.
This observation is consistent, for example, with a primarily flat heterointerface without steps, though by itself it is not sufficient to characterize the interface quality.  
Nonetheless these labels can be ambiguous at high energies where multiple valley-orbit excitations are present. 
This is particularly so for P5, where only one ``orbital-like'' excited state with fast decay is observed, accompanied by twin ``valley-like'' states with slower decay rates. 
To reiterate, only a single valley-like excitation per orbital is expected in our SiGe devices with large built-in biaxial tensile strain in the Si layer. 
The dual slow peaks could be because another state overlaps or otherwise hybridizes with the visible levels, or possibly because it lies beyond the probed energy range.
Indeed, one of those twin peaks could actually be a manifestation of the 2nd excited orbital, with its decay rate observed, as expected, to be highly suppressed, likely by symmetry, relative to the 1st excited orbital; this latter pattern is consistent with the structure of the higher excited peaks observed in the P6 spectrum in Fig. 3(b)-(d)-(f).
Alternatively, a slower 2nd excited orbital peak in P5 could be masked by a much faster decay into the close-in-energy 1st excited orbital. 
This possibility seems less plausible in view of the comparable decay rates of the accompanying twin peaks and orbital semi-degeneracy which would indicate an accidental approximate axial symmetry of the in-plane confinement. 
Note the substantially differing orbital splittings of P6 imply asymmetry in the confining potential (see \cite{Zwanenburg2013} Section E.2 and \cite{HRLJones2019} for an in-depth discussion of the confinement symmetry effects on the dot energy spectrum).

\begin{table*}[tp]	
\setlength{\tabcolsep}{12pt}
\caption{
Extracted peak energies and assigned valley splittings for different orbital excitations for dots P5 and P6, based on Lorentzian fits to the data in Fig.~\ref{fig3}.
All energies are in units of $\upmu$eV. 
Asterisk for selected P5 data denotes the uncertain interpretation of those peaks, as discussed in the text.}
\label{tab:DotP5P6}
\begin{center}
\begin{tabular}{c|ccc|ccc}
	\hhline{=======}
	 & \multicolumn{3}{c|}{P5} & \multicolumn{3}{c}{P6} \\
	Orbital & Ground & Excited & Valley & Ground & Excited & Valley \\
	Assignment & Valley Peak & Valley Peak & Splitting & Valley Peak & Valley Peak & Splitting \\
	\hline
	Ground & $2 \pm 7$ & $-258 \pm 9$ & $260 \pm 11$ & $-1 \pm 8$ & $161 \pm 6$ & $162 \pm 10$ \\
	1st Excited & $-1184 \pm 14$ & $-1445 \pm 12$ & $261 \pm 14$ & $1,393 \pm 12$ & $1615 \pm 22$ & $222 \pm 23$ \\
	2nd Excited & N/A$^*$ & $-1520 \pm 13$ & $336 \pm 15$$^*$ & $2125 \pm 23$ & $2275 \pm 26$ & $150 \pm 27$ \\
	\hhline{=======}
\end{tabular}
\end{center}
\end{table*}

\begin{figure}[bp]
\includegraphics[width=0.8\linewidth]{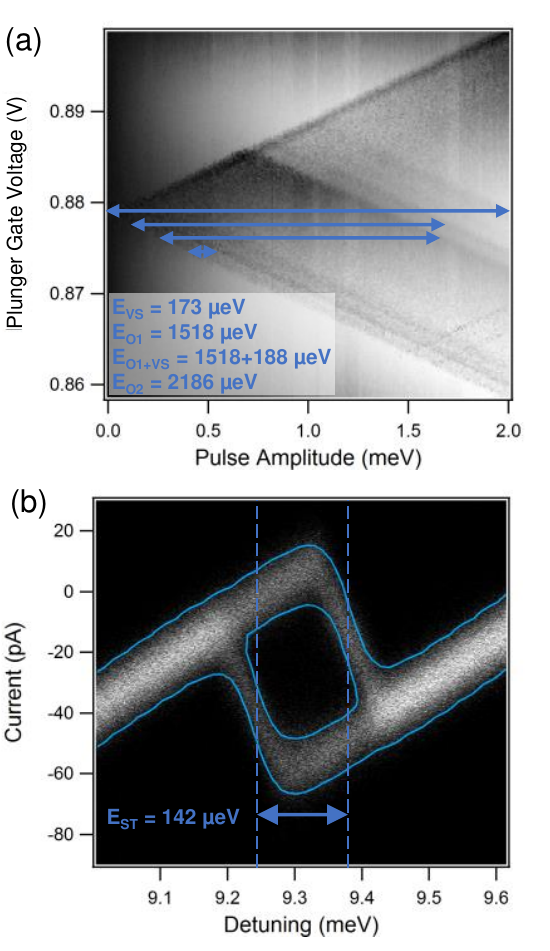}
\caption{
\textbf{(a)} Pulsed-gate spectroscopy of P6 in a voltage configuration similar to that in Fig.~\ref{fig3}\textbf{(b)}. The resulting energies (e.g., $E_{O1/2}$ denotes the first/second orbital excitation) compare favorably to those in Table~\ref{tab:DotP5P6}, supporting our assignments. 
\textbf{(b)} Single-shot measurements of (1,1)-(0,2) transition for P6, showing two-electron singlet-triplet splitting ($E_{ST}$) from spin-blockade spectroscopy.
}
\label{figA3}
\twocolumngrid
\end{figure}

The precise detuning location of the transition peaks can vary depending on the charge dynamics, as discussed in App. \ref{app:ForRevDAPS}; however, these shifts are typically small in scale so that energy splittings (peak separations) can be reliably extracted. 
As one test of consistency, we find good agreement between multiple excited state energies for P6 taken from DAPS and those obtained using pulsed-gate spectroscopy \cite{elzerman2004excited} as shown in Fig.~\ref{figA3}(a). 
In contrast to DAPS, the latter measurement requires direct loading from an electron reservoir and hence cannot be done on inner dots (like P5) without substantial retuning.
The P6 valley splitting inferred from DAPS (162~$\upmu$eV) and pulsed-gate (173~$\upmu$eV) measurements is also close to the largest singlet-triplet splitting of 142~$\upmu$eV measured using two-electron spin blockade spectroscopy~\cite{HRLJones2019}, as shown in Fig.~\ref{figA3}(b); this suggests the latter is valley-limited in this bias tune-up.

The DAPS method is less susceptible to extraneous resonant tunneling features because electron motion between the dots is not directly coupled to leads. 
Similar spectroscopic features are consistently seen in many dots (as in Fig. 3), suggesting these are typically true dot features. 
We have tried to further suppress coupling with baths by reverse biasing the corresponding barrier gates during DAPS; this doesn't change the peak features, further confirming they are signs of dot spectroscopy rather than bath loading/unloading. 
Spurious signals due to fluctuating charges in other (especially accidentally formed) QDs and charge traps are always a possibility, even despite tightly controlling most of the device surface potential with biased gates, but would be expected to also cause artefacts in charge stability diagrams and other tuning experiments, which we generally do not observe for the dots measured here.

\begin{figure}[tp]
\includegraphics[width=0.9\linewidth]{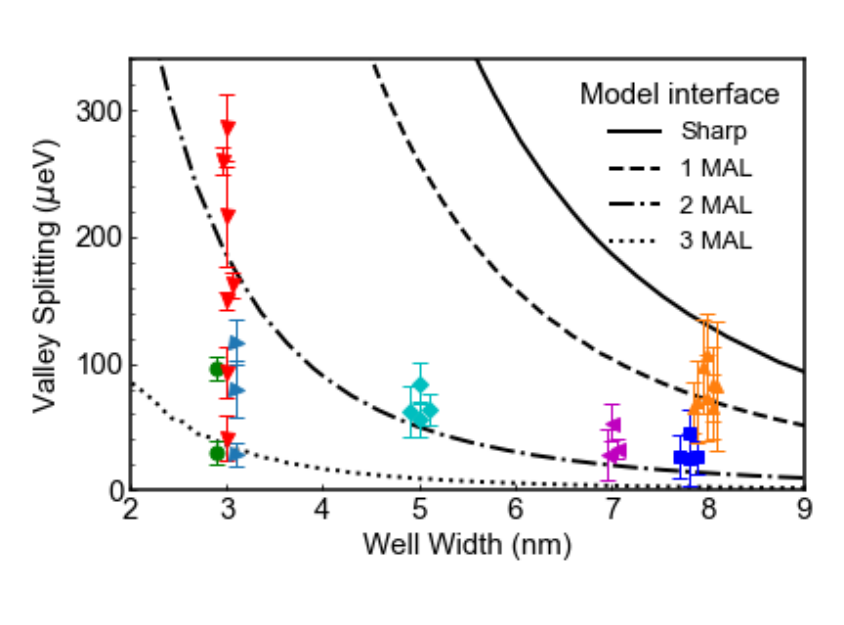}
\caption{
Ground orbital valley splittings of 31~QDs from devices across seven wafers differentiated by symbol color and shape. 
The Si well widths span 3, 5, 7, and 8 nm; QDs with similar VS at the same well width are slightly displaced for visual clarity. 
Error bars are extracted from half-width half-maximum (HWHM) of Lorentzian fits of the extracted peaks similar to Figs.~\ref{fig2} and ~\ref{fig3}.
Black curves correspond to tight-binding predictions of valley splitting versus well width as a function of front interface sharpness in units of mono-atomic-layer (MAL), as described in the text and App.~\ref{app:vsmodel}.
}
\label{fig4}
\twocolumngrid
\end{figure}

With this general understanding and confidence in the DAPS technique, we proceed to characterize \textit{en masse} QDs in several devices fabricated on wafers grown with target QW widths from 3 to 8~nm. 
The QW widths are confirmed to be close to their targets by X-ray reflectometry. 
With this larger data set of extracted energies from DAPS, we can examine the reproducibility of measured values across devices from different wafers, dots within each device, excited states within each dot, and variations with heterostructure parameters.
Fig.~\ref{fig4} compiles the ground orbital valley splitting from 31~QDs, grouped by wafer. 
The largest valley splitting here of $286\pm26$~$\upmu$eV is on par with the highest previously reported in the literature for Si/SiGe QDs (which was in a depletion-mode device)~\cite{borselli2011measurement}.

For comparison, we also plot the dependence of valley splitting on well width and interface sharpness as predicted by empirical full-band tight-binding calculations~\cite{niquet2009onsite}. 
These calculations assume an unbiased QW, as we expect a small out-of-plane electric field in our devices, and a smooth back interface, as obtained from previous characterization studies~\cite{dyck2017accurate}. 
In general, values fall below the predictions for the atomically sharp interface, suggesting some modest degree of interfacial broadening. 
While interface sharpness is a convenient modeling parameter, many other factors strongly affect the predicted VS, as discussed in more detail in App.~\ref{app:vsmodel}. 
As a result, the exact sharpness of experimental heterostructures cannot be quantified from the theoretical curves in Fig.~\ref{fig4}; the latter are only intended to illustrate the qualitative importance of interface quality for VS. 
The tendency of measurements from the same wafer to cluster, especially for wider wells, suggests that differences between epitaxial wafers contribute to the observed variation.

Inter-dot variations in valley splitting are most apparent for the 3~nm well devices, suggestive of increased overlap of the electron wave function in narrower wells with a microscopically inhomogeneous interface. 
This is consistent with the observation shown in App.~\ref{app:vsbias} that the valley splitting can sometimes vary substantially as the electron is translated along the QW via electrostatic bias \cite{hollmann2019large}. 
Another consequence of this intra-dot disorder is that the valley splitting of different orbitals within the same dot also tends to vary, as suggested by Table~\ref{tab:DotP5P6} and further supported by measurements on multiple dots. 

\begin{figure}[tp]
\includegraphics[width=0.9\linewidth]{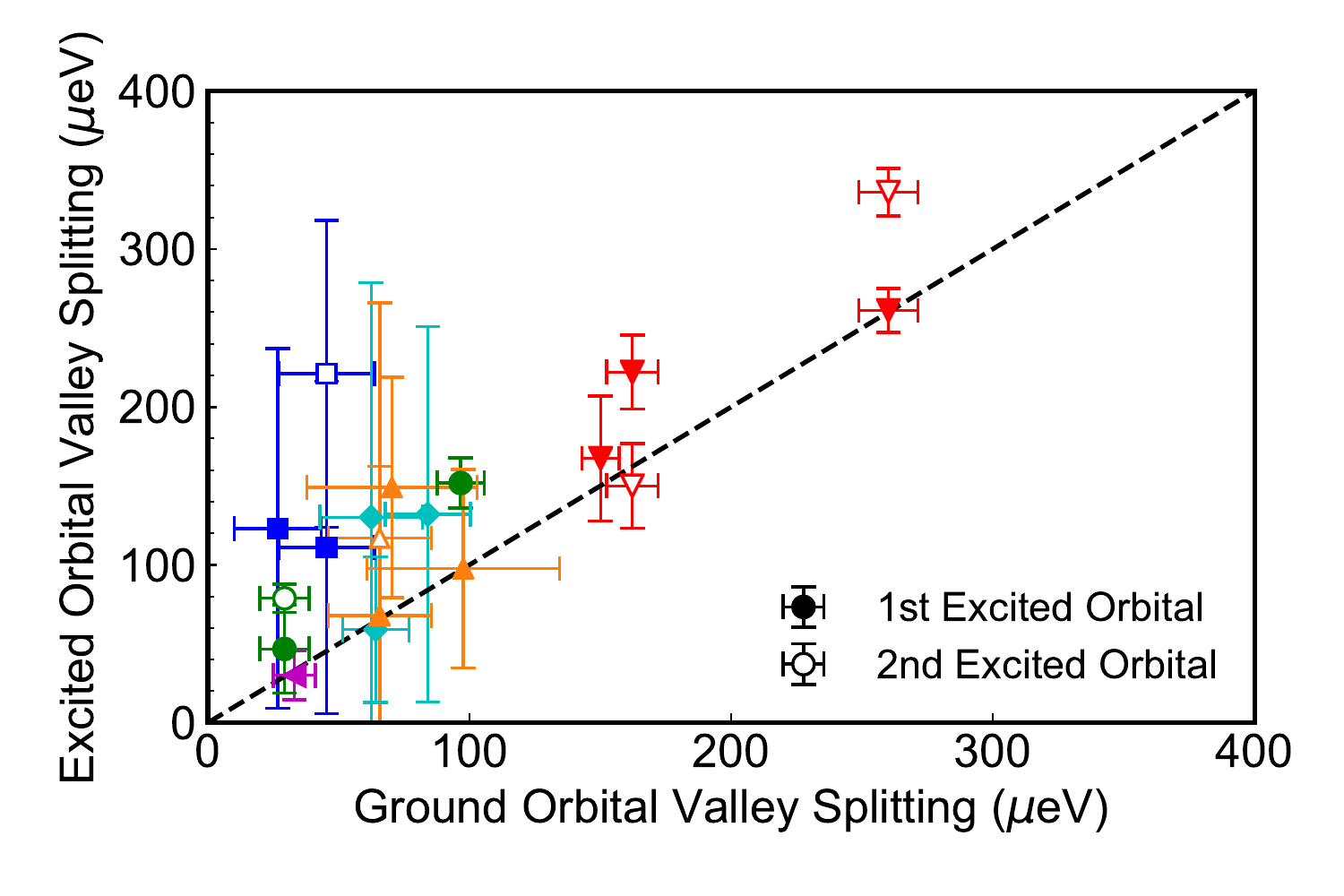}
\caption{
Measurements of valley splitting in the ground, first, and second orbital states from six devices from five different wafers where each symbol color and shape corresponds same devices as in Fig.~\ref{fig4}\textbf{(a)}. 
Error bars are from Lorentzian fits of peak widths.
Valley splittings of both the ground and first excited orbital states (solid) were measured on fifteen dots, while valley splittings of the second orbital states were measured on five dots (empty). 
The diagonal line is included for visual reference to indicate variations in valley splitting between different orbitals.
}
\label{figA10}
\twocolumngrid
\end{figure}

In Fig.~\ref{figA10}, we observe a correlation of valley splitting between ground and excited orbitals, which is notably higher than between different dots in the same device (Fig.~\ref{fig4}). 
This observation is highly suggestive of greater interface homogeneity on length scales akin to the spatial extent of single electrons ($\approx$30 nm from modeling) compared to that of the distance between dots ($\approx$150 nm). 
Curiously, in most cases the excited orbital valley splitting values are larger, particularly when the ground valley splitting is small. 
We attribute this to systematic bias in resolving high energy peaks. 
As noted in Fig.~\ref{fig2}(c), excited orbitals generally relax quickly and hence have faster inter-dot transition rates, broadening the associated DAPS peaks and making it difficult to identify small splittings in the vicinity of multiple orbital excitations.

\section{Conclusions}\label{main:conclusions}
Using the DAPS technique, we can straightforwardly extract multiple excited state energies in single-electron quantum dots. 
From our measurements, we infer distinct ranges of valley (10s-300 $\mu$eV) and orbital splittings (1-2 meV) which are well-separated in energy and generally consistent with expectations based on theory, simulations, and prior experimental observations. 
Auxiliary information, such as the variation of charge decay rates for different transitions, further corroborates our valley-orbit assignments for these states. 
Definitive interpretation for higher excited states is difficult in some cases where multiple valley and orbital states are similar in energy, as we have taken care to clearly illustrate.

The relative simplicity of the DAPS method allows probing of valley splitting disorder within and between QDs as well as different wafers, which will be critical for identifying and improving optimal epitaxial conditions for Si/SiGe qubits. 
The largest splittings in our results were observed in 3~nm well devices, though further work is needed to fully understand all other sources of variation. 
We note that the well widths explored here are narrower than typically used for Si/SiGe qubits in other works (where they frequently span 8-18~nm), but have found this does not prevent the formation or hinder the performance of spin qubits~\cite{HRLAndrews2019,sun2020full,blumoff2020quantifying}. 
We have also explored the dependence of magnetic dephasing on quantum well width in a separate work~\cite{kerckhoff2020magnetic}.

Although we present single-electron measurements in this work, the DAPS technique can be extended to study multi-electron state spectra since it relies on the general effects of charge dephasing and relaxation, such as the two-electron spectra extraction discussed in App.~\ref{app:spindaps}. 
Our approach may also be applicable for studying the rich interplay of the orbital, valley, and spin degrees of freedom in quantum dot states in other systems besides Si/SiGe, such as Si MOS~\cite{gamble2016valley}, III-V semiconductors~\cite{petta2004}, or two-dimensional materials~\cite{wang2018electrical}.

\newpage
\begin{acknowledgments}
The authors thank Sieu Ha, Sean Meenehan, Seth Merkel, and Emily Pritchett for valuable discussions.
\end{acknowledgments}

\bibliography{qdotref}
\appendix
\renewcommand\thefigure{A\arabic{figure}} 
\setcounter{figure}{0} 
\section{DAPS with RF Transitions}\label{app:rfdaps}

\begin{figure}[tp]
	\includegraphics[width=0.9\linewidth]{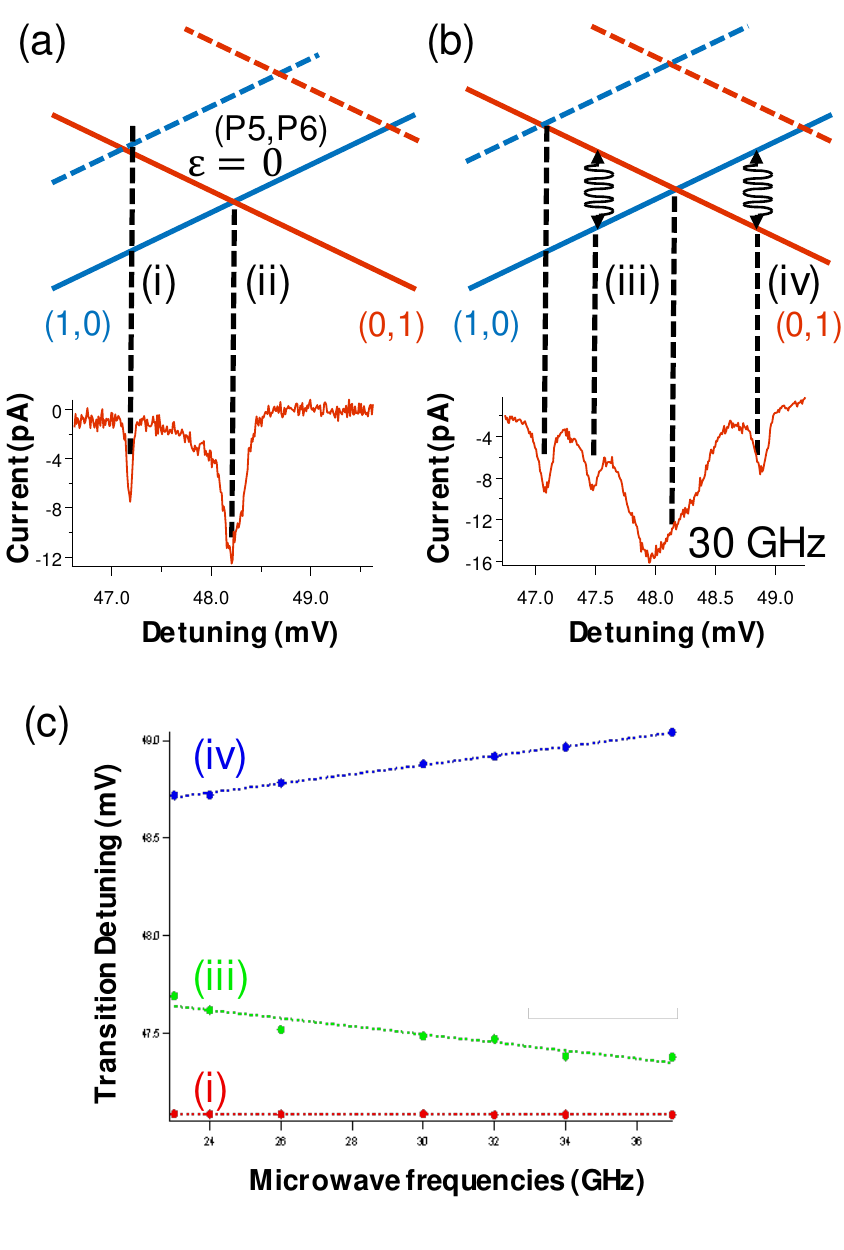}
	\caption{
		\textbf{(a)} DAPS spectrum of QD at a hold time of 5 us without applying continuous microwave drive. 
		Transition (i) is the ground-to-excited transition, while (ii) is the ground-to-ground transition.
		\textbf{(b)} Same experimental pulse sequence at a hold time of 10 us, with 30 GHz continuous microwave drive applied through a neighboring gate. 
		Two new transitions, labeled as (iii) and (iv), appear due to photon-assisted tunneling. 
		\textbf{(c)} Tracking transitions (i), (iii) and (iv) versus microwave frequency. 
		The average of the slopes (blue and green) result in a detuning scale factor of 0.186 eV/V.
	}
	\label{figA4}
	\twocolumngrid
\end{figure}

\begin{figure}[tp]
	\includegraphics[width=0.8\linewidth]{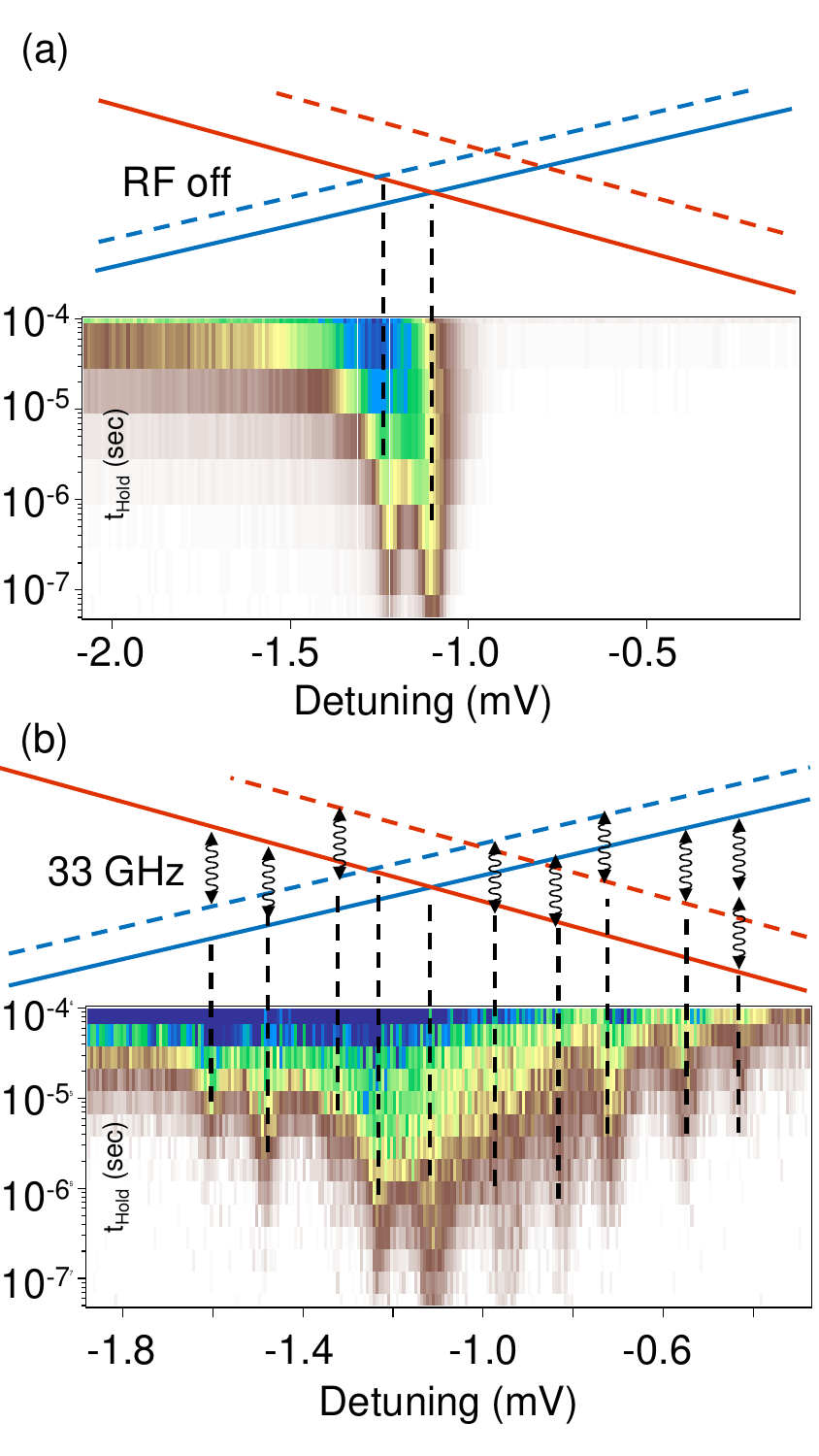}
	\caption{
		\textbf{(a)} DAPS measurement at small detunings for a QD without RF drive. 
		This experimental procedure is identical to those described in the main text.
		\textbf{(b)} DAPS measurement for same QD but with 33 GHz RF tone applied to a nearby gate. 
		Some two-photon transitions are also visible, whose positions correspond well with the DC and RF resonances expected from the energy level diagram depicted. 
		The faint transition on far right potentially corresponds to a two-photon resonance.
	}
	\label{figA5}
	\twocolumngrid
\end{figure}

Converting applied detuning voltages into energy scales is commonly done by synthesizing separate measurements of gate lever arms and cross-capacitances~\cite{HRLJones2019}. 
This approach, although expedient, is prone to systematic errors. 
We address this concern by operating the DAPS sequences exactly as described in the main text while also applying a continuous-wave microwave tone to drive photon-assisted tunneling (PAT) transitions. 
These experiments reveal PAT transitions in the DAPS spectra (Fig.~\ref{figA4}(b)). 
By sweeping the frequency of the applied tone (Fig.~\ref{figA4}(c)) and fitting a linear slope to the two frequency-dependent transitions, we are able to more accurately convert detuning voltages into an energy scale by using the slope as the so-called detuning scale factor (DSF), which has units of eV per V. 
Sweeping both detuning and hold time with an applied CW tone reveals even more PAT transitions whose spacings are consistent with our original interpretation of the transitions as excited states in the dots, as seen in Fig.~\ref{figA5}. 
The visibility of these measurements varies significantly in practice depending on the sensitivity of the QD to applied microwave tones.

\section{Model of DAPS Dynamics}\label{app:model}
The main dynamic features of our experiments are captured by a standard Lindblad master equation analysis. 
When a DQD is detuned in the immediate vicinity of an inter-dot transition, the relevant parameters are the detuning from the anti-crossing ($\varepsilon$), the inter-dot tunnel coupling ($t_c$), the charge dephasing rate ($\kappa$), and inelastic decay ($\Gamma$, which is detuning dependent), so that the dynamics of the two levels of the anticrossing can be described by the evolution of the density matrix
\begin{equation}\label{eq:lindblad2}
\dot{\rho}(t) = -i[\varepsilon \sigma_z + t_c \sigma_x,\rho] - \frac{\kappa}{2}(\sigma_z \rho \sigma_z - \rho) + \mathcal{L}_{\Gamma(\varepsilon)}(\rho).
\end{equation}
It is straightforward to show that when $\kappa > t_c$ and $\Gamma$ is neglected, the system decays to a 50-50 mixed charge state at a relaxation rate approximately given by $\dfrac{\kappa t_c^2}{\kappa^2 + \varepsilon^2}$, motivating the use of a Lorentzian fit of the decay rate around each transition for determining the peak position. 
This yields a peak in decay rate at the anti-crossing, which is the central feature we seek; however, away from the anti-crossing, inelastic decay is necessary to thermalize the final state population to equilibrium. 
We describe the latter with the decay superoperator in Eq. \ref{eq:lindblad2}, where the inelastic decay rate $\Gamma$ can arise from couplings to a variety of environmental degrees of freedom, including evanescent wave Johnson noise (EWJN) from metal gates and neighboring 2DEG reservoirs~\cite{langsjoen2012qubit} and acoustic phonons in Si~\cite{tahan2014relaxation}. 
While the precise magnitudes of the lifetimes depend sensitively on device features, we expect that at small detuning, Johnson noise-induced relaxation scales as $\Gamma_{JN} \propto t_c^2 / \varepsilon$ while phonon-induced relaxation scales as $\Gamma_{ph} \propto t_c^2 \varepsilon^3$. 
Importantly, this suggests that at small and moderate DQD detuning, the inelastic decay rate may be slower than the dephasing-induced charge decay near excited state anti-crossings, allowing the latter to be resolved. 
At larger detuning, phonon-induced decay dominates and its magnitude must include the competing influence of phonon bottleneck effects, increased phonon density of states, and enhancement of the transition matrix elements due to the spatially extended wave functions of excited states. 
Numerical calculations show strong enhancement of inelastic decay in this regime, which can make transitions at excited anti-crossings harder to pick out.

\begin{figure}[tp]
\includegraphics[width=0.9\linewidth]{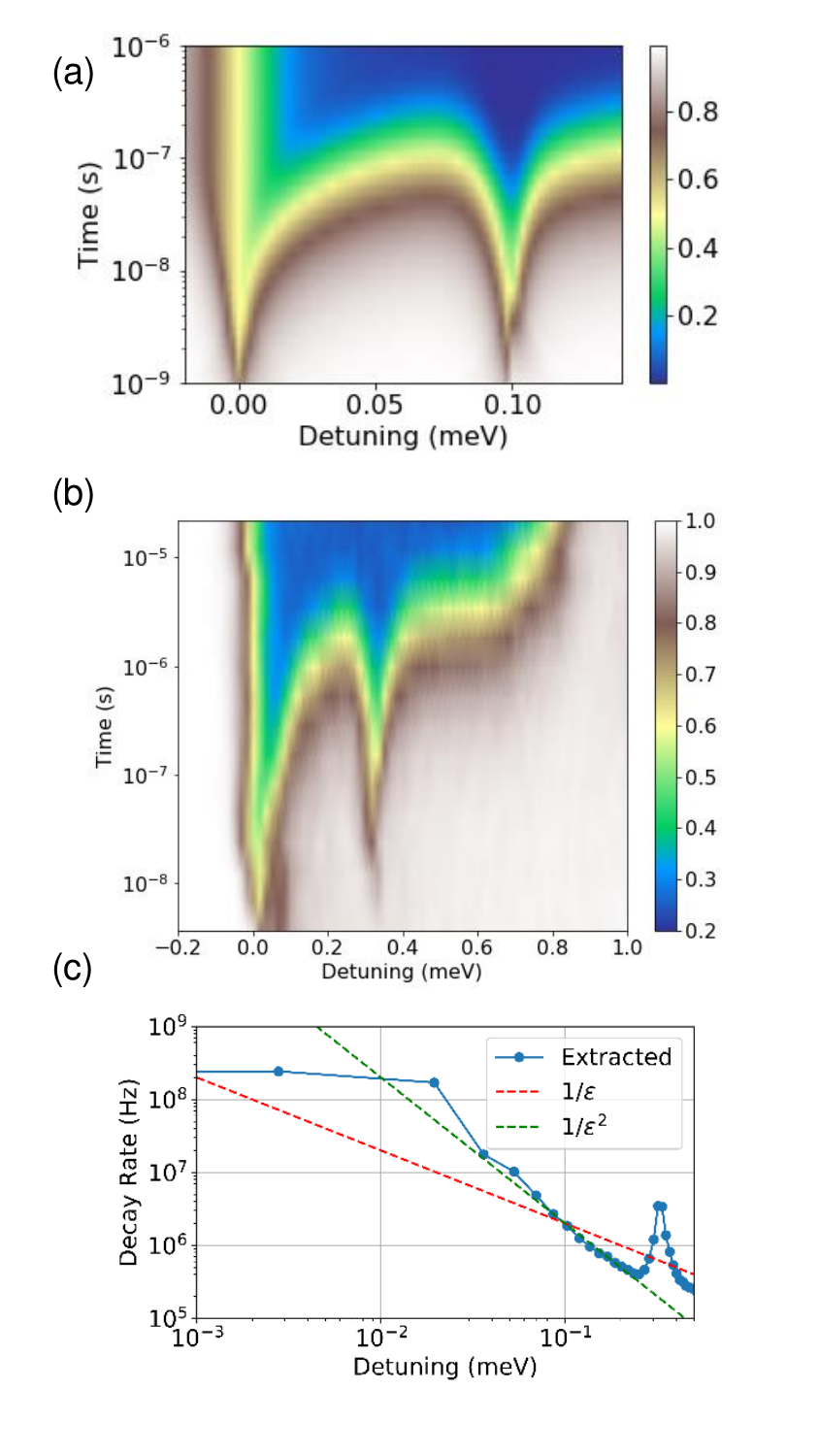}
\caption{
\textbf{(a)} Theoretically calculated decay of initial (1,0) charge state as a function of detuning and hold time in a three-level model including ground and valley (0,1) states.
\textbf{(b)} DAPS decay measurement for one QD as function of detuning and hold time in the vicinity of the ground orbital and valley transitions.
\textbf{(c)} Extracted charge decay rate from data \textbf{(b)} as function of detuning; dashed lines are fits to $1/\varepsilon$ and $1/\varepsilon^2$ dependences showing the latter better matches the experimental dependence at low energies.
}
\label{figA2}
\twocolumngrid
\end{figure}

To show how these dynamics are relevant to resolving excited state energies, we extend this model to a three-level calculation of dynamics as a function of detuning and hold time as shown in Fig.~\ref{figA2}(a), modeling the DAPS decay of the charge state of one dot into the ground and valley excited states of another dot. 
The calculations qualitatively reproduce the experimental decay dynamics seen at low detuning in Fig.~\ref{figA2}(b) as well as Fig.~3(a)-(b) of the main text. 
One important feature of both the model and experimental results, as discussed in ~App.~\ref{app:ForRevDAPS}, is that the locations corresponding to inter-dot peaks tend to shift towards positive detuning as the hold time increases; this occurs because charge relaxation $\Gamma$ tends to asymmetrize the charge state population by pushing it towards thermalization. 
As hold time increases, the detuning location of each peak will therefore be offset from the true anti-crossing position, which will usually be located on the rising shoulder of the peak. 
Therefore, peak positions should ideally be extracted at short hold times for the most accurate extraction of excited state energies.

Interestingly, the extracted inelastic decay rates shown in Fig.~\ref{figA2}(c) scale as $\varepsilon^{-2}$ at small detuning. 
One possible reason for this discrepancy between model and experiment is that some other decay mechanism besides phonons or thermal Johnson noise dominates at very low energies, as other experiments in Si/SiGe QDs have also suggested~\cite{wang2013charge,hollmann2019large}. 
While this model captures the essential features of the DAPS technique, it neglects important details such as the presence of $1/f$ charge noise, whose effects are not simply parameterized by a Markovian dephasing rate $\kappa$ and may contribute to non-exponential decay and the broadening of transition peaks in experiments. 
Recently, a similar model has been applied to describe the dynamics during readout in singlet-triplet qubits~\cite{seedhouse2020parity}.

\section{Peak Alignments for a DQD Pair}\label{app:ForRevDAPS}
\begin{figure}[tp]
	\includegraphics[width=0.9\linewidth]{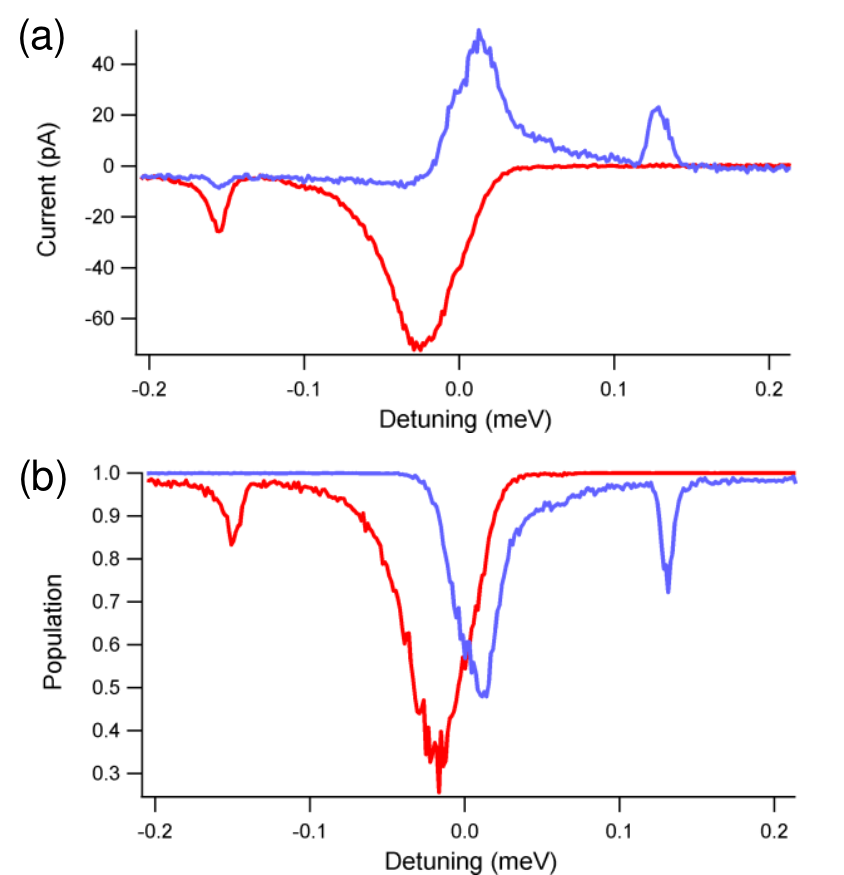}
	\caption{
		\textbf{(a)} The P5 (red) and P6 (blue) DAPS sequences at two different hold times (6 and 0.5~$\upmu$s) have opposite charge signals due to their opposite charge state references. 
		Comparing the two reveals a detuning offset between the maxima of the ground-ground transition peaks. 
		\textbf{(b)} P6 and P5 DAPS measurement sequence as in \textbf{(a)} except the lock-in signal is measured in terms of relative population between the reference charge state and the charge state post-dephasing. 
		A value of 1.0 represents 100\% of the population remaining in the initial charge state after dephasing at the detuning point (x-axis), while 0.0 represents 0\% of the initial population remaining in the initial charge state. 
	}
	\label{figA1}
	\twocolumngrid
\end{figure}
The DAPS technique works on a pair of adjacent dots, and irrespective of which dot the reference measurement is performed on, there is only one ground-ground anti-crossing for the dot pair. 
Therefore when DAPS measurements for both dots are directly compared, the expectation is the ground-ground peak locations in detuning voltage should exactly align. 
In practice, however, we find that the maxima of the peaks may not be quite aligned with each other as shown in Fig.~\ref{figA1}(a). 
This detuning offset between the peaks persists even when the measurement is taken in terms of the relative populations between the two charge states, Fig.~\ref{figA1}(b), and suggests the location of the anti-crossing is between the onset of the peak and the maximum of the peak. 
This observation can be explained by asymmetry versus detuning of inelastic charge thermalization $\Gamma$ and is reproduced by modeling, as discussed in App.~\ref{app:model}, and can also be influenced by non-ideal pulses. 
It motivates our decision to label the uncertainty for each DAPS peak as approximately the half-width at half-maximum (HWHM) obtained from Lorentzian fits. 
This error can be alleviated by operating at shorter hold times to suppress inelastic decay, at the cost of reduced signal.

\section{Valley Splitting Modeling}\label{app:vsmodel}
We perform empirical $spds^*$ tight-binding calculations of valley splitting to understand the theoretically expected dependence on well width. 
Valley splitting models for Si/SiGe QWs frequently assume perfectly abrupt interfaces, which may be interrupted by steps or tilts, as reviewed for example in~\cite{Zwanenburg2013}. 
In the case of an electron confined by two perfectly abrupt interfaces, the valley splitting oscillates as the QW width changes at the mono-atomic-layer (MAL) scale due to inter-valley phase interference~\cite{friesen2007valley,nestoklon2008electric,zhang2013genetic}. 
In general, however, growth kinetics and atomic diffusion during epitaxy and device fabrication will broaden the interfaces, as confirmed by microscopy and atom probe tomography~\cite{dyck2017accurate}. 
These measurements suggest that the back interface is typically broader than the front interface, which would suppress valley interferometric effects as a function of well width. 
Additional asymmetries, such as the presence of a strong electric field, would similarly depress valley oscillations.

Our tight-binding calculations use the parameters for bulk Si and Ge in\cite{niquet2009onsite}, with the Si$_{0.7}$Ge$_{0.3}$ layers described within the virtual crystal approximation (VCA). 
We model each QW interface using a sigmoidal function for the Ge concentration:
\begin{equation}\label{eqn:chi}
	\chi_{\pm}(x)=\frac{\chi_{\text{SiGe}}}{1+e^{(x_{0,\pm}\pm x)/\tau_{\pm}}}
\end{equation}
where $x_{0,\pm}$ denote interface position and $\tau_{\pm}$ is the interface width parameter for the front and back interfaces, respectively;  $4\tau$ is the distance over which the Ge percentage varies from 12\% to 88\% of its value in the barrier region $\chi_{SiGe}$. 
We fix $4\tau_{-}$ for the back interface at 8 MAL, following the extracted values in \cite{dyck2017accurate}, and vary the sharpness of the front interface $4\tau_{+}$ between 0-3 MAL to obtain the results shown in Fig.~\ref{fig4}. 
Zero applied vertical electric field is assumed, since electrostatic simulations of our devices indicate a negligible field under typical operating conditions. 
In general, a finite positive electric field will tend to increase the valley splitting as the electron overlaps more strongly with the sharp front interface; this effect is more prominent for wider well widths.

While the $\tau_+$ of the front interface is the only free parameter in our calculations, in practice a large number of additional factors will affect the predicted VS values. 
Our calculations are for 1-D QW structures and thus do not include disorder effects beyond the homogeneous smoothing of the interfaces; such disorder would further broaden the expected VS distribution~\cite{jiang2012effects}. 
Additionally, while the empirical tight-binding parameters for Si and Ge are obtained by fitting to bulk band structure features, multiple parametrizations are possible and the relevant inter-valley coupling is not uniquely constrained at present. 
In our experience, commonly used Si and Ge $spds^*$ tight-binding parameter sets in the literature~\cite{boykin2004valleyB,jancu2007tetragonal,sacconi2004full} can lead to variations of up to 2X in predicted VS. 
Furthermore, any true heterostructure is ultimately composed of substitutional Ge atoms in the barrier regions and therefore also includes local strain. 
3-D atomistic calculations including these effects can lead to differences from 1-D VCA models~(as discussed in the supplemental section of \cite{Zwanenburg2013}, for example). 
Because of the quantitative importance of and uncertainty in these and other factors, these calculations are only intended to serve as a qualitative indicator of the sensitivity of VS to interface effects, not an attempt to quantify interface sharpness $\tau$ using measured VS values.

\section{Valley Splitting Dependence on Electrostatic Bias}\label{app:vsbias}
\begin{figure}[tp]
\includegraphics[width=0.9\linewidth]{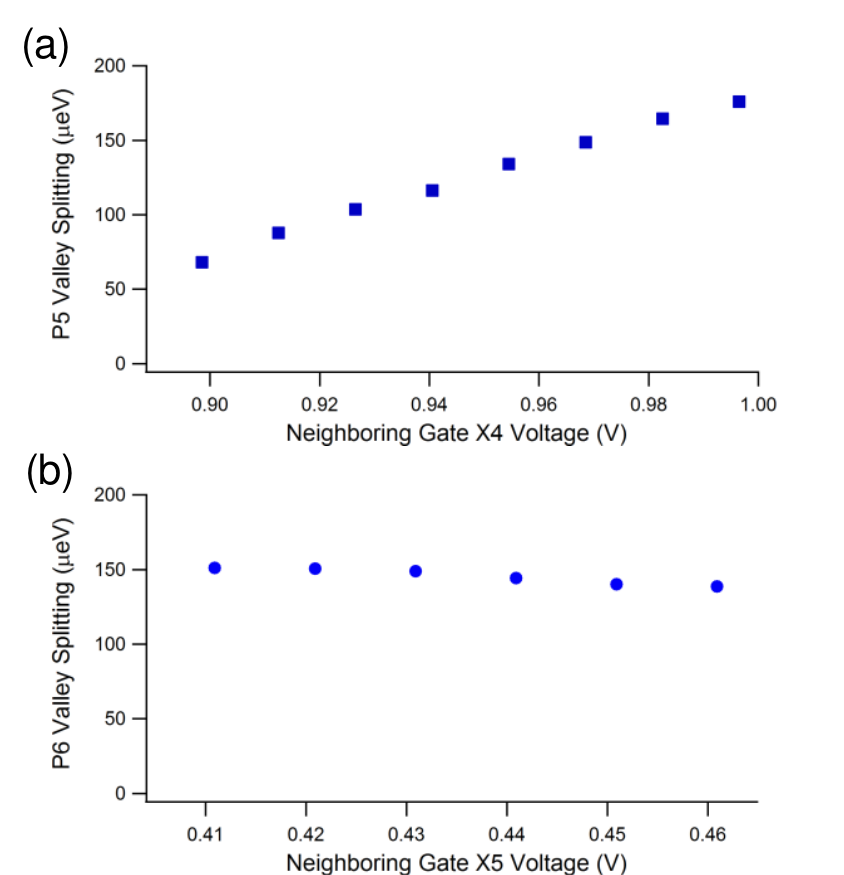}
\caption{
Variation of valley splitting on either side of a P5-P6 DQD for a \textbf{(a)} P5 electron as function of bias on X4 and \textbf{(b)} P6 electron as function of bias on X5 gate.
}
\label{figA8}
\twocolumngrid
\end{figure}

To observe the sensitivity of energy splittings on the electrostatic confinement, we can evaluate how much DAPS transitions are affected by changing the bias of a nearby gate (typically a neighboring tunneling barrier), which alters the convexity of the potential confinement along one of the two spatial directions and causes lateral displacement of the quantum dot along that same direction. 
In the ideal scenario, we expect orbital-like transitions to be more sensitive to these changes than valley-like transitions. 
However, the valley mixing can also experience substantial changes as the electron is translated along a microscopically varying heterointerface, due to some combination of disorder effects such as interfacial steps and/or intrinsic atomistic random alloy fluctuations. 
In Fig. \ref{figA8} we observe that two neighboring dots (P5 and P6 in a device) exhibit very different behavior in the measured VS as a function of bias on different adjacent gates; P5 shows significant change (by about a factor of 3X within the probed bias range) while its neighboring dot retains a roughly constant VS. 
This illustrates both the importance of local microscopic disorder on the VS as well as the possibility that such disorder can be uncorrelated between different dots. 
The trends we detect in bias tunability of VS are consistent with previous observations in the literature~\cite{Shi2013,hollmann2019large}.

\section{Two-Electron DAPS Spectroscopy}\label{app:spindaps}
While one-electron spectra are most transparent for understanding dot confinement, valley splitting, and decoherence pathways of single spins in dots, it is the energies of two-electron eigenstates that affect fidelities of the spin initialization and readout in spin qubit proposals. 
The DAPS technique can be readily applied to this case, and many useful variants can be envisioned. In particular, the singlet and triplet spectra of a two-electron QD can be selectively probed in different ways using either a DQD or triple QD (TQD) device.

A simple DQD experiment could initialize a (2,0) ground singlet, adiabatically separate it to the (1,1) state, and then diabatically pulse along the detuning axis scanning for resonances with two-electron states of either one dot--–when going towards (2,0)---or the other (towards (0,2)).
Due to spin conservation, this would predominantly probe singlets. 
This pulse sequence could be modified by adding a wait time $t_{dephasing}$ after adiabatic separation of the singlet. 
When $t_{dephasing}$ is comparable to the singlet-triplet dephasing lifetime $T_2^*$, both singlets and triplets should become accessible during the scan along the detuning axis. 
However, crowding of states and conflicting optimal $t_{\text{hold}}$ values for different states can complicate extraction of the energy levels.

\begin{figure}[tp]
\includegraphics[width=0.75\linewidth]{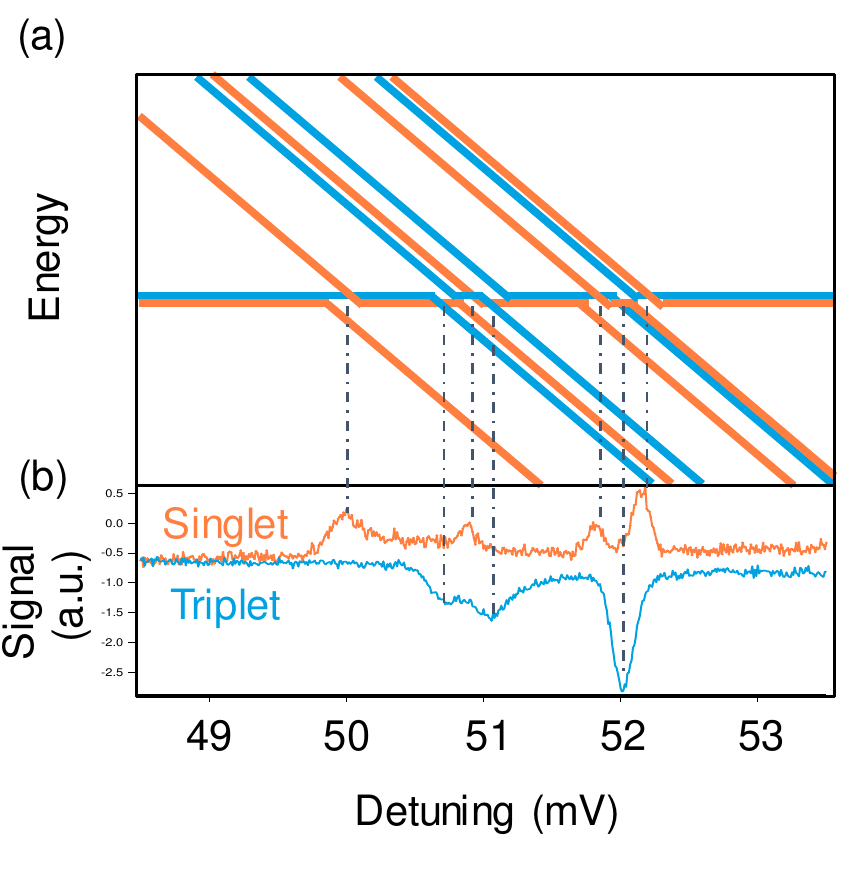}
\caption{
\textbf{(a)} Schematics of 2-electron eigenenergies near (1,1)-(2,0) charge transition, allowing for valleys and excited orbitals.
\textbf{(b)} Measured spin-dependent spectra of 2-electron states. Singlets in orange, triplets in blue.
}
\label{figA9}
\twocolumngrid
\end{figure}

Alternatively, with a third dot we can form an exchange-only qubit~\cite{divincenzo2000,HRLAndrews2019} that we can electrically manipulate to separately map out the two-electron singlet and triplet eigenstates of a singlet dot, as shown in Fig.~\ref{figA9}. 
The sequence of steps for these experiments is very similar to that outlined above.  
As an example, the dots underneath plungers (P6,P5,P4) are operated around the (2,0,1)-(1,1,1) charge transition. 
A spin singlet is first initialized in the P6 dot and then adiabatically ramped into the (1,1,1) charge cell. 
From this point, diabatic detuning pulses towards P6 result in two-electron eigenstates of the P6 dot (or P5 if going in the opposite direction along detuning axis) coming into and out of resonance with the P6/P5 (1,1) singlet state. 
Dephasing at the anti-crossing points followed by a diabatic pulse back into the (1,1,1) charge configuration results in a charge state-dependent outcome. 
The charge state measurement is subtracted from the result in a second round of re-initialization followed by a charge state measurement. 
The resulting lock-in measurement for increasingly larger detuning pulses towards the P6 dot results in a series of peaks where the (1,1) singlet level crosses with the two-electron, spin-singlet eigenstates in P6. 
This measurement would be the same as in a DQD experiment, resulting in the orange line in Fig.~\ref{figA9}(b) with pronounced singlet peaks. 
Additionally, in a TQD qubit a composite $X_{\pi}$ Clifford gate using exchange pulses can be inserted after initialization to prepare a entangled three-spin state which projects only onto (1,1) P6/P5 triplet states. 
Subsequent diabatic detuning pulses into P6 then reveal only the resonances between the (1,1) triplets with two-electron spin-triplet eigenstates in P6 (Fig.~\ref{figA9}(b), blue line). 
The spin-singlet and spin-triplet spectra of the P6 dot can be compared to determine whether valley or orbital states are limiting the lowest singlet-triplet energy splitting~\cite{dassarma2010}, and to assess the degree of valley-orbit coupling present in the dot.

\end{document}